\documentclass[twoside]{article}
\usepackage{fleqn,espcrc2}
\usepackage{graphicx}

\newcommand{\AmS}{{\protect\the\textfont2
  A\kern-.1667em\lower.5ex\hbox{M}\kern-.125emS}}

\usepackage{latexsym}

\usepackage[T1]{fontenc}

\usepackage[latin1]{inputenc}

\usepackage{graphicx}
\usepackage{color,pst-plot}

\usepackage{amsmath}
\usepackage {amsfonts,amssymb,amstext}
  
\newcommand{\lbl}[1]{\label{#1}}
\newcommand{ \rf}[1]{(\ref{#1})}

\newcommand{\be}{\begin{equation}}
\newcommand{\ee}{\end{equation}}{
\newcommand{\bea}{\begin{eqnarray}}
\newcommand{\eea}{\end{eqnarray}}
\newcommand{\setl}{\setlength\arraycolsep{2pt}}

\newcommand{\noi}{\noindent}
\newcommand{\nn}{\nonumber}
\newcommand{\ra}{\rightarrow}

\newcommand{\cO}{{\cal O}}

\newcommand{\Imm}{\mbox{\rm Im}}

\newcommand{\MeV}{\mbox{\rm MeV}}
\newcommand{\GeV}{\mbox{\rm GeV}}

\newcommand{\GF}{G_{\mbox{\rm {\tiny F}}}}

\title{Present Status of the Muon Anomalous Magnetic Moment}

\author{Eduardo de Rafael\address{Centre de Physique Th\'eorique, 
        CNRS-Luminy, Case 907, F-13288 Marseille Cedex 9, France}%
        \thanks{Unit\'e Mixte de Recherche (UMR 6207) du CNRS et des Universit\'es Aix-Marseille~1, Aix-Marseille~2 et sud Toulon-Var, affili\'ee \`a la FRUMAN. This work has been supported in part by the European Community's Marie Curie Research Training Network program under contract No. MRTN-CT-2006-035482, Flavianet.}}
        
\begin{document}

\begin{abstract}
These pages, based on my talk  at the Montpellier {\it 14th International Conference in QCD},   provide us with a short update of the Standard Model contributions to the muon anomalous magnetic moment.
\end{abstract}

\maketitle

\section{Introduction}

We shall be concerned with the $g$--factor of the muon which relates its spin to its magnetic moment 
\begin{equation}
\vec{\mu}=g_{\mu}\frac{e\hbar}{2m_{\mu}c}\vec{s}\,,\qquad
\underbrace{g_{\mu}=2}_{\mbox{\rm \small
Dirac}}(1+a_{\mu})\,;
\end{equation}
more precisely with the correction $a_{\mu}$ to the Dirac value $g_{\mu}\!=\! 2$ which generates  the so called anomalous magnetic moment. The present experimental world average determination, which is dominated by the latest BNL experiment (the E821 collaboration \cite{Benetal06}), is
\begin{equation}\lbl{WA}
{a_{\mu}^{\rm exp}\!=\!116~592~080(63)\times 10^{-11}} ( 0.54~\mbox{\rm ppm} )\,,
\end{equation}
where the origin of the error is $0.46~{\rm ppm}$ statistical and $0.28~{\rm ppm}$ systematic. This determination assumes CPT--invariance i.e., $a_{\mu^-}=a_{\mu^+}$. 

The question we shall discuss is: {\it how well can the Standard Model digest this precise number?}. As we shall see, the precision of $a_{\mu}^{\rm exp}$ is such that it is sensitive to the three couplings of the Gauge Theory which defines the Standard Model, as well as to its full particle content~\footnote{ For a recent  review article see e.g. ref.~\cite{MdeRR07}.}.

\section{The QED Contributions (Leptons)}

This is by far the dominant contribution, which is generated by two types of Feynman diagrams: 

\begin{table*}
\setlength{\tabcolsep}{1.5pc}
\newlength{\digitwidth} \settowidth{\digitwidth}{\rm 0}
\catcode`?=\active \def?{\kern\digitwidth}
\caption{QED Contributions (Leptons)  \{$\alpha^{-1}=137.035~999~084~(51)~[0.37~{\rm ppb}]$~\cite{Gabetal06}\} }
\label{tab:QED}
\begin{tabular*}{\textwidth}{@{}l@{\extracolsep{\fill}}rrr}
\hline \hline {\sc\small Contribution} &
{\sc\small  Result in Powers of $\frac{\alpha}{\pi}^{~}$} & {\sc\small Numerical Value in $10^{-11}$ Units} 
\\ \hline \hline  
$a^{(2)}$ & $0.500~000~000~(00)\left(\frac{\alpha}{\pi}\right)\ $ & $116~140~973.29~(0.04)$\\ 
\hline
$a^{(4)}$   & $-0.328~478~965~(00)\left(\frac{\alpha}{\pi}\right)^2$ & \\ 
$a_{\mu}^{(4)}(\rm total)$ & $0.765~857~410 ~(27)\left(\frac{\alpha}{\pi}\right)^2$ & 413~217.62~(0.01)\\ 
\hline
$a^{(6)}$   & $1.18~124~146~(00)\left(\frac{\alpha}{\pi}\right)^3$ & \\
$a_{\mu}^{(6)}(\rm total)$ & $24.05~050~964~(43)\left(\frac{\alpha}{\pi}\right)^3$ & 30~141.90 (0.00)\\ 
\hline
$a^{(8)}$ & $-1.9~144~(35)\left(\frac{\alpha}{\pi}\right)^4$ & \\
$a_{\mu}^{(8)}(\rm total)$  & $130.8~055~(80)\left(\frac{\alpha}{\pi}\right)^4$ &
381.33~(0.02) \\ 
\hline
$a_{\mu}^{(10)}(\rm total~estimate)$ & $663~(20)\left(\frac{\alpha}{\pi}\right)^5$ & 4.48~(1.35)  \\ 
\hline
 $a_{\mu}^{(2+4+6+8+10  )}(\rm QED)$ &  &
  $116~584~718.09~( 0.14 )(0.04)$  \\
 \hline\hline

\end{tabular*}
\end{table*}

\subsection{The Massless Class}
This   class is generated by Feynman diagrams with virtual photons only as well as by  diagrams with virtual photons and  fermion loops of the same flavour as the external particle (the muon in our case). Since $a_{\mu}$ is a dimensionless quantity, this class of diagrams gives rise to the same contribution to the muon, electron and tau anomalies. They correspond to the entries $a^{(2n)}$ in Table~1, with $n=1,2,3,4$ indicating the number of loops involved. They are known analytically at one loop~\cite{Sch49}; two loops~\cite{Pe57,So57}; and three loops~\cite{LaRe96}. This is the reason why there is no error in the corresponding numbers in the second column of Table~1. 

At the four--loop
level, there are 891 Feynman diagrams of this type. Some of them are already
known analytically, but in general one has to resort to numerical methods for
a complete evaluation. This impressive calculation, which is systematically pursued by Kinoshita and collaborators, requires many
technical skills and  
 is under constant updating; in particular thanks to the  advances 
in computing technology. The entry $a^{(8)}$ in Table~1 is the one corresponding to the most recent published value~\cite{Kino07},
with the error due to the present numerical uncertainties.

Notice the alternating sign of the results from the 
contributions of one loop to four loops, a simple feature which is not yet
{\it a priori} understood. Also, the fact that the sizes of the
$\left(\frac{\alpha}{\pi}\right)^n$ 
coefficients for $n=1,2,3,4$  remain rather small is an interesting feature, 
 allowing one to expect that the order of magnitude of the five--loop
 contribution, from a total of $12~672$ Feynman diagrams, is
 likely to be of  $\cO\left (\alpha/\pi\right)^5\simeq 7\times 10^{-14}$.
This is
 well beyond the accuracy required to compare with the present 
experimental result for $a_{\mu}$, but it will be eventually needed 
for a more precise determination of the fine--structure
 constant $\alpha$ from the precision measurement of the electron anomaly~\cite{Gabetal06}.
 
\subsection{The Massive Class}
This second class is generated by Feynman diagrams with lepton loops of a different flavour to the one of the external muon line. Their contribution to $a_{\mu}$ is then a function of the lepton mass ratios involved. These contributions are generated by vacuum polarization subgraphs and by   light--by--light scattering subgraphs involving electron and tau loops. Both the two--loop and three--loop contributions of this class are known analytically~\footnote{ For a history of the successive improvements in the evaluation of these contributions see e.g.  ref.~\cite{MdeRR07}.}. The full three--loop evaluation involving electron--loop subgraphs, by Laporta and Remiddi ~\cite{L93,LR93}, is a remarquable achievement. The numerical errors quoted in Table~\ref{tab:QED} for these contributions are due to the present experimental errors in the lepton masses~\cite{Pa05}.

At the four--loop level, only a few contributions  are known analytically. Kinoshita and his collaborators have, however,  accomplished a full numerical evaluation of this class (see ref.~\cite{KN04} and references therein.). The corresponding error in Table~\ref{tab:QED} is the combined error in the lepton masses and the present error due to the numerical integration. 

The number quoted for the full five--loop QED contribution in Table~\ref{tab:QED} is the present estimate quoted in ref.~\cite{KN06}. It is likely to be improved in the near future.   

\subsection{The Mellin--Barnes Technique}

There has been a recent technical development in the evaluation of Feynman diagrams in\-vol\-ving mass ratios, which has already been useful in the evaluation of some higher order contributions to $a_{\mu}$ (see refs.~\cite{FGdeR05,AGdeR08}) and which seems promising for further calculations. In these papers it is shown how the Mellin--Barnes integral representation of Feynman parametric integrals allows for an easy evaluation of as many terms as wanted in the asymptotic expansion of Feynman diagrams in terms of one and two mass ratios.

The basic idea is to express the contribution to $a_{\mu}$ from a Feynman diagram, or a class of diagrams, as an inverse Mellin transform with respect to the mass ratios involved in the diagrams. 
The remarkable property of this representation is the factorization in terms of moment integrals. It is in fact this factorization which is at the basis of the renormalization group properties  discussed in ref.~\cite{LdeR74}, and used since then by many authors (see e.g. ref.~\cite{Kat06} and references therein). The algebraic factorization in the Mellin--Barnes re\-pre\-sen\-ta\-tion, however, is more general. The standard renormalization group constraints only apply to the evaluation of asymptotic behaviours in terms of  {\it powers of logarithms} and constant terms. In the Mellin--Barnes framework, this appears as a property of the residue of the leading Mellin singularity.  What is new here is that this extends as well to the subleading terms, which are governed by the residues of the successive Mellin singularities (in the negative real axis, in the case of electron loops);  or by two--dimensional residue forms~\cite{AGdeR08,A08}, in the case of the Mellin singularities associated to two mass ratios (the case of both electron and tau loops).   

As an example, we quote a few terms of the result obtained for the tenth--order contribution from the string of vacuum polarization subgraphs shown in Fig.~1:

{\footnotesize
{\setl
\bea\lbl{Aeeetau}
a_{\mu}^{(eee\tau)} & = & \left(\frac{\alpha}{\pi} \right)^5 \left\{\left(\frac{m_\mu^2}{m_\tau^2}\right)\left[\frac{4}{1215}\log^3\frac{m_\mu^2}{m_{e}^2}\right.\right.  
\nn \\  & &  -\frac{2}{405}\log^2 \frac{m_\mu^2}{m_{e}^2} -\left(\frac{122}{3645}-\frac{8\pi^2}{1215}\right)\log\frac{m_\mu^2}{m_{e}^2}
   \nn \\  & & \left. \left.
 +\frac{2269}{32805}-\frac{4\pi^2}{215}-\frac{16}{405}\zeta(3)\right]
 + \cdots \right\}\nn \\   
& =  & \left(\frac{\alpha}{\pi} \right)^5 0.013~057~4(4)\,.    
\eea}}

\begin{figure}

\begin{center}
\includegraphics[width=0.3\textwidth]{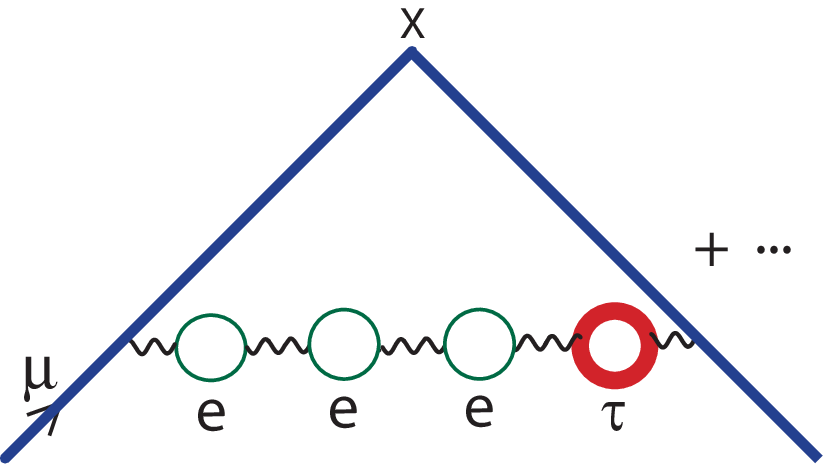}

{\bf Fig.1}~{\it\small  Diagrams with three e--loops and a $\tau$--loop.}

\end{center}
\vspace*{-1cm}
\end{figure}
\noi
In fact, the analytic calculation in ref.~\cite{AGdeR08} which leads to this precise number, also includes terms up to $\cO\left[\left(\frac{m_{\mu}^2}{m_{\tau}^2}\right)^4\log^3 \frac{m_\mu^2}{m_\tau^2}\right]$, which are already smaller than the error generated by the lepton masses in the leading order terms.

\begin{figure}

\begin{center}
\includegraphics[width=0.2\textwidth]{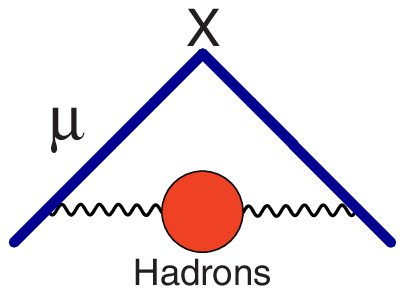}

{\bf Fig.2}~{\it\small Hadronic Vacuum Polarization}\label{figure:hvp}

\end{center}
\vspace*{-1cm}
\end{figure}

\section{Hadronic Contributions}

The electromagnetic interactions of hadrons produce contributions to $a_{\mu}$ induced by the hadronic vacuum polarization and by the hadronic light--by--light scattering. 

\subsection{Hadronic Vacuum Polarization}

All calculations of the lowest--order hadronic vacuum polarization
contribution to the muon anomaly (see Fig.~2) are based on 
 the spectral 
representation~\cite{BM61}

{\footnotesize 
\begin{equation}\lbl{lohvp}
        a_{\mu}^{\rm hvp}= \frac{\alpha}{\pi}\int_0^{\infty}\frac{dt}{t}\frac{1}{\pi}
        \Imm\Pi (t)\! \int_0^1 \!\! dx\frac{x^2 (1-x)}{x^2 + \frac{t}{m_{\mu}^2}(1-x)}
\end{equation}}

\noi
with the hadronic spectral function $\frac{1}{\pi}
        \Imm\Pi (t)$  related to the {\it one-photon} $e^+ e^-$ annihilation cross-section into hadrons ($m_e \ra 0$) as follows:
        
{\footnotesize         
\begin{equation}\lbl{onephoton}
        \sigma(t)_{\{e^+ e^- \ra (\gamma)\ra {\rm hadrons}\}}=\frac{4\pi^2\alpha}{t} \frac{1}{\pi}
        \Imm\Pi (t)\,.
\end{equation}}

\noi
This contribution is dominated by the $\pi^+ \pi^-$ channel; the region of the $\rho$--resonance in par\-ti\-cu\-lar~\cite{GdeR69,BdeR69}. The history of evaluations of $a_{\mu}^{\rm hvp}$ is a long one which can be traced back, e.g. in ref.~\cite{MdeRR07}. The most recent compilation of $e^+ e^-$ annihilation   data used in the evaluation of the dispersive integral in Eq.~\rf{lohvp} made by Michel Davier and collaborators, which also includes the new precise measurements from the experiments SND and CMD-2 at Nobosibirsk as well as some exclusive channels from BaBar, gives the result~\footnote{See e.g. ref~\cite{ZZ08} and references therein for details.}:
\be\lbl{exphvp}
a_{\mu}^{\rm hvp}=(6~908\pm 39_{\exp} \pm 19_{\rm rad}\pm 7_{\rm QCD})\times 10^{-11}\,.
\ee
Unfortunately, the discrepancy with the evaluation made using the $\tau$--spectral functions, corrected for isospin--breaking effects, still persists. Here, one has to wait for the forthcoming results from the high precision measurements on the $\pi\pi$ mode at BaBar using the radiative return method. We shall then be able to check the consistency with the result in Eq.~\rf{exphvp} and, hopefully, improve the accuracy.

There is a similar spectral representation to the one in Eq.~\rf{lohvp} for the  next--to--leading order  hadronic vacuum polarization~\cite{CNPdeR76}, with the kernel~\cite{BdeR68,LdeR68} in Eq.~\rf{lohvp}, replaced by a two--loop kernel, which is also  known analytically~\cite{BR75}. The most recent numerical evaluation, using the same data as for the lowest--order evaluation, gives
\be\lbl{exphvpnlo}
a_{\mu}^{\rm hvp(nlo)}=(-97.9\pm 0.9_{\exp} \pm 0.3_{\rm rad}\ )\times 10^{-11}\,.
\ee

\subsection{Hadronic Light--by--Light Scattering}

Unlike  the hadronic vacuum polarization  contribution, there is no direct experimental input for the hadronic light--by--light scattering contribution to $a_{\mu}$ shown in Fig.~3; therefore one has to rely
on theoretical approaches.

So far, the only rigorous theoretical result is the
observation that, in the QCD large--$N_c$ limit  and to leading order
in the chiral expansion, the dominant contribution  comes from the Goldstone--like neutral pion exchange which produces a cha\-rac\-te\-ris\-tic universal double logarithmic behavior with a coefficient which can be calculated exactly~\cite{KNPdeR02}:

{\footnotesize
\be
\lbl{piefft}
a^{\rm hll}_{\mu}(\pi^{0})\!=\!\Big(\frac{\alpha}{\pi}\Big)^{3}\frac{m_{\mu}^{2}N_{c}^2}{48\pi^2 F_{\pi}^{2}}
\Big[ \ln^2\!\frac{m_{\rho}}{m_{\pi}}  +
\cO\Big(\!\ln\!\frac{m_{\rho}}{m_{\pi}}\!\Big)+\cO(\!1\!)\Big]
\ee}

\noi
where $F_{\pi}$ denotes the pion coupling constant in the chiral limit
($F_{\pi}\sim 90~\MeV$).
Testing this limit was particularly useful in fixing the sign of the phenomenological calculations of the neutral pion exchange~\cite{KN02}.

Although the coefficient of the $\ln^2 (m_{\rho}/m_{\pi})$ term in Eq.\,\rf{piefft} is unambiguous, the coefficient
of the $\ln(m_{\rho}/m_{\pi})$ term depends on low--energy constants which are difficult to extract from experiment~\cite{KNPdeR02,RMW02}  (they require a detailed know\-led\-ge of the $\pi^0 \ra e^+ e^-$ decay rate with inclusion of radiative corrections).
Moreover,  the constant term in Eq.~\rf{piefft} is not fixed by chiral symmetry requirements, which makes the predictive power of an effective  chiral perturbation theory approach rather limited for our purposes. Therefore, one has to adopt a dynamical framework which takes into account explicitly the heavier meson degrees of freedom as well.

The most recent calculations of $a^{\rm hll}_{\mu}$ in the literature~\cite{KN02,HK02,BPP02,MV04} are all compatible with the QCD chiral constraints and large--$N_c$ limit discussed above. They all incorporate the $\pi^0$--exchange contribution modulated by $\pi^0 \gamma^* \gamma^*$  form factors, correctly normalized to the $\pi^{0}\to \gamma\gamma$ decay width. They differ, however, in the shape of the form factors, originating in different assumptions: vector meson dominance  in a specific form of Hidden Gauge Symmetry in Ref.~\cite{HK02};  in the form of the extended Nambu--Jona-Lasinio (ENJL) model  in Ref.~\cite{BPP02}; large--$N_{c}$ models in Refs.~\cite{KN02,MV04}; and on whether or not they satisfy the particular operator product expansion constraint  discussed in Ref.~\cite{MV04}.

\begin{figure}

\begin{center}
\includegraphics[width=0.2\textwidth]{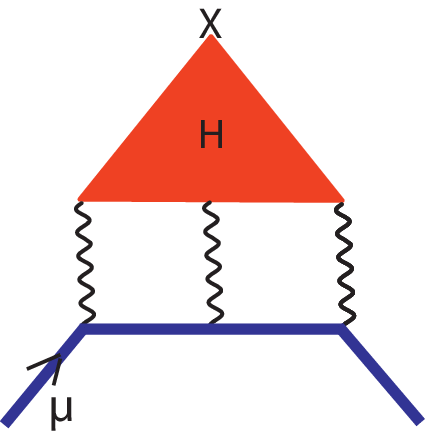}

{\bf Fig.3}~{\it\small  Hadronic Light--by--Light Scattering}

\end{center}
\vspace*{-1cm}
\end{figure}

In order to compare  different
results it is convenient to separate the hadronic light--by--light contributions  which are leading in 
the $1/N_c$--expansion from the non-leading ones \cite{deR94}. 
Among the leading contributions, the pseudoscalar meson exchanges which incorporate the $\pi^0$, and to a lesser degree the $\eta$ and $\eta'$ exchanges, are the dominant ones. As discussed above,  there are good QCD theoretical reasons for that. In spite of the different definitions of the pseudoscalar meson exchanges and the associated choices of the  form factors used in the various model calculations, there is a reasonable agreement among the final results. The result quoted in a recent update discussed in ref.~\cite{PdeRV08} gives:
\be
a^{\rm hll}(\pi\,,\eta\,,\eta')=(114\pm 13)\times 10^{-11}\,.
\ee
Other contributions, which are also leading in the $1/N_c$--expansion, due to axial--vector exchanges and scalar exchanges, give smaller contributions with updated errors, as discussed in ref.~\cite{PdeRV08}:
\be
a^{\rm hll}(1^+)=(15\pm 10)\times 10^{-11}\,,
\ee
and 
\be
a^{\rm hll}(0^+)=-(7\pm 7)\times 10^{-11}\,.
\ee

The subleading contributions in the $1/N_c$--expansion are dominated by the charged pion loop. However, because of  the model dependence of the results one obtains when the pion loop is dressed with  hadronic interactions it is suggested in ref.~\cite{PdeRV08} to use the central value of the ENJL--model evaluation in~\cite{BPP02}, but with a larger error which also covers  unaccounted loops of
other mesons, :
\be
a^{\rm hll}(\pi^+ \pi^-)=-(19\pm 19)\times 10^{-11}\,.
\ee

From these considerations, adding the errors in quadrature, as well as the small charm contribution: $a^{\rm hll}(c)=2.3\pm\times 10^{-11}$ , one gets
\be
a^{\rm hll}=(105\pm 26)\times 10^{-11}\,,
\ee
as a final estimate.

\section{Electroweak Contributions}

The leading contribution to $a_{\mu}$ from the
Electroweak Lagrangian of the Standard Model, 
o\-ri\-gi\-na\-tes at the one--loop level. The relevant Feynman diagrams (in the unitary gauge) are shown in Fig.~4.
The analytic evaluation of the overall contribution 
gives the result (see e.g. ref.~\cite{BGL72}):
 
{\setl
\footnotesize
\bea\lbl{EW1}
a_{\mu}^{\mbox{\rm\tiny
 EW(1)}} &\!\! =\!\! &
\frac{G_{\mbox\tiny F}}{\sqrt{2}}\frac{m_{\mu}^2}{8\pi^2}
\left\{\underbrace{\frac{10}{3}}_{W} + 
\underbrace{\frac{1}{3}(1\!-\!4\sin^{2}\theta_{W})^2-\frac{5}{3}}_{Z}  \right.
\nn
\\
&  + & \!\!\left. \cO\!\left(\!
\frac{m_{\mu}^2}{M_{Z}^2}\log\frac{M_{Z}^2}{m_{\mu}^2}\!\right)
\!+\!\frac{m_{\mu}^2}{M_{H}^2}
\int_{0}^{1}\!\! dx 
\frac{2x^2
(2-x)}{1-x+{\frac{m_{\mu}^2}{M_{H}^2}}x^2}
\!\right\} \nn  \\
 & = & 194.8\times 10^{-11}\,, 
\eea}

\noi
where the weak mixing angle is defined by $\sin^2 \theta_{W}\!
=\! 1-M^2_W/M_Z^2\simeq0.223$,
and $G_F\simeq1.166\times 10^{-5}$ is the Fermi constant.
Notice that the contribution from the Higgs boson, shown
in parametric form in the second line, is of $\cO\left(\frac{G_{\mbox\tiny F}}{\sqrt{2}}\frac{m_{\mu}^2}{4\pi^2}\frac{m_{\mu}^2}{M_H^2}\ln\frac{M_H^2}{m_{\mu}^2}\right)$, rather
small for the present lower bound on $M_H$, but known analytically.

The {\it a priori} possibility that the two--loop
electroweak corrections may bring in enhancement factors due to large
logarithms, like $\ln(M_Z^2 /m_{\mu}^2)\simeq 13.5$, has motivated a
thorough theoretical effort for their evaluation, which has been quite a
remarkable 
achievement. 

\begin{figure}

\begin{center}
\includegraphics[width=0.45\textwidth] {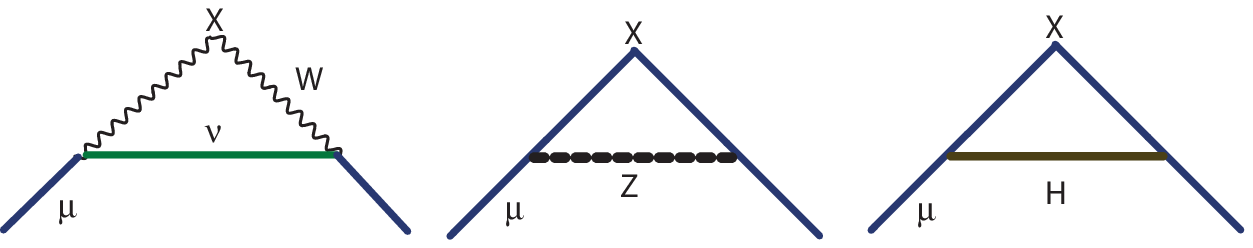}

{\bf Fig.4}~{\it\small  Weak interactions at the one-loop level}

\end{center}
\vspace*{-1cm}
\end{figure}

It is convenient to separate the two--loop electroweak contributions into
two sets: those containing closed fermion loops 
and the bosonic corrections, which we denote by 
$a_{\mu}^{EW(2)}({\mbox{\rm\footnotesize bos}})$. 
The latter have been evaluated using asymptotic 
techniques in a systematic expansion in powers of
$\sin^2\theta_{W}$, where $\log\frac{M_{W}^2}{m_{\mu}^2}$
terms, $\log\frac{M_{H}^2}{M_{W}^2}$ terms, $\frac{M_{W}^2}{M_{H}^2}
\log\frac{M_{H}^2}{M_{W}^2}$ terms, $\frac{M_{W}^2}{M_{H}^2}$ terms,
and constant terms are kept. Using 
$\sin^2\theta_{W}=0.223$ and $50~\GeV\le M_H \le 700~\GeV$ results in~\cite{CKM95,HSW04,GC05}:

{\footnotesize
{\setl
\bea
a_{\mu}^{EW(2)}({\rm bos}) &  =  & \frac{G_F}{\sqrt{2}}
\frac{m_{\mu}^2}{8\pi^2}\times \frac{\alpha}{\pi} (-82.2\pm 5.9)\nn \\
& = & (-22.2 \pm 1.6)
\times 10^{-11}\,.
\eea}}

The discussion of the fermionic corrections is
more delicate. Because of the $U(1)$ anomaly
cancellation between lepton loops and quark loops in the electroweak theory, one cannot separate
hadronic from leptonic effects any longer  in  diagrams like the ones shown
in Fig.~5, where a VVA--triangle with two vector currents and an 
axial--vector current appears. 
It is therefore appropriate to separate
the fermionic corrections into two subclasses. One is
the class in Fig.~5, which   we
denote by
$a_{\mu}^{EW(2)}(l,q)\,.$ The other class is defined by the rest of the diagrams, where quark
loops and lepton loops can be treated separately, which we call
$a_{\mu}^{EW(2)}({\mbox{\rm\footnotesize ferm-rest}})$.
\begin{figure}

\begin{center}
\includegraphics[width=0.4\textwidth]{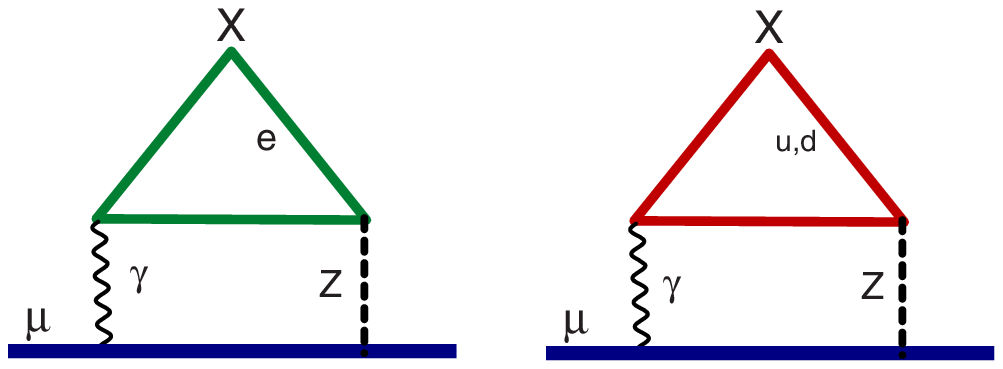}

\end{center}

{\bf Fig.5}~{\it \small Two-loop electroweak diagrams generated by the
$\gamma\gamma Z$-Triangle. There are similar diagrams corresponding to the $\mu, c,s$ and $\tau, t,b$ generations. }\label{figure:WA}

\vspace*{-0.75cm}

\end{figure}
This latter contribution has been
estimated to a very good approximation in ref.~\cite{CKM95} with the
result,

{\footnotesize
\be\lbl{ferrest}
a_{\mu}^{EW(2)}({\mbox{\rm\footnotesize
ferm-rest}})=\frac{\GF}{\sqrt{2}}\,\frac{m_{\mu}^2}{8\pi^2}\,
\frac{\alpha}{\pi}\times\left(-21\,\pm\,4 \right)\,,
\ee}

\noi
where the error here is the one induced by diagrams with Higgs 
propagators with an allowed Higgs mass in the
range $114~\GeV \!< \! M_H \!<\! 250~\GeV$.

Concerning the contributions to
$a_{\mu}^{EW(2)}(l,q)$, it is convenient to treat
the three generations separately. The
contribution from the third generation can be calculated in a
straightforward way  using effective field theory techniques~\cite{PPdeR95}, because all the fermion masses in the triangle loop are large with respect to the muon mass, with the result~\cite{PPdeR95,CKM95}:

{\footnotesize
\be
 a_{\mu}^{EW(2)}  (\tau,t,b)  = \frac{G_F}{\sqrt{2}}\frac{m_{\mu}^2}
{8\pi^2}\,
\frac{\alpha}{\pi}\times  (-30.6)\,.
\ee}

\noi
However,
as first emphasized in ref.~\cite{PPdeR95}, an appropriate QCD calculation
when the quark in the loop of Fig.~5 is a {\it light quark} should
take into account the dominant effects of spontaneous chiral-symmetry
breaking. Since this involves the $u\,,d$ and $s$ quarks, it is convenient to lump together the
contributions  from the first and second generations. An
 evaluation of these contributions, which
incorporates the QCD long--distance chiral realization~\cite{PPdeR95,KPPdeR02} as well as perturbative~\cite{CMV03} and non--perturbative~\cite{KPPdeR02,CMV03} 
short--distance constraints, gives the result

{\footnotesize
\be\lbl{12gs}
a_{\mu}^{EW(2)}(e,\mu,u,d,s,c)=\frac{\GF}{\sqrt{2}}\,\frac{m_{\mu}^2}
{8\pi^2}\,
\frac{\alpha}{\pi} \times (-24.6\pm 1.8)\,.
\ee}

\noi
From the theoretical point of view, this calculation has revealed
surprising  properties concerning the {\it non-anomalous} component of the
VVA--triangle~\cite{Vain03},
resulting in a new set of {\it non-renormalization theorems} in perturbation theory~\cite{Vain03,KPPdeR04}.
   
Putting together the partial two--loop results  discussed above, one finally obtains for the overall electroweak contribution the
value

{\footnotesize
{\setl
\bea
a_{\mu}^{\rm EW} & = & a_{\mu}^{\rm EW(1)} +\frac{G_F}{\sqrt{2}}\frac{m_{\mu}^2}
{8\pi^2}\left(
\frac{\alpha}{\pi}\right)  [-158.4(7.1)(1.8)]\nn \\ & = &
  152(2)(1)
\times 10^{-11}\,, 
\eea}}

\noi
where the first error is essentially due to the Higgs mass
 uncertainty, while the second comes from
 hadronic uncertainties in the VVA--loop evaluation. The overall
 result shows indeed that the two--loop correction represents
 a sizeable reduction
of the one--loop result by an amount of $22\%$. An evaluation of the electroweak three--loop
leading terms of $\cO\left[\frac{\GF}{\sqrt{2}}\,\frac{m_{\mu}^2}
{8\pi^2}\left(\frac{\alpha}{\pi}\right)^2 \ln\frac{M_Z}{m_{\mu}}\right]$,
 using renormalization group arguments~\cite{DG98,CMV03}, shows that higher order effects
 are negligible [$\cO(10^{-12})$] for the accuracy needed at present. 

\section{Summary}

Table 2 collects the various Standard Model contributions to $a_{\mu}$ which we have discussed. Notice that the largest error at present is the one from the lowest order hadronic vacuum polarization contribution. Adding experimental and theoretical errors in quadrature gives a total 
\be
a_{\mu}^{\rm SM}= (116~591~785 \pm 51)\times 10^{-11}\,,
\ee
with an overall error slightly smaller than the one in the experimental determination in Eq.~\rf{WA}.
The comparison between these two numbers, shows an intriguing $3.6~\sigma$ discrepancy.

\begin{table}
\ \  Table~2 ~Standard Model Contributions
\begin{center}

{\footnotesize
\begin{tabular}{lr}
\hline \hline {\sc\small Contribution} &
{\sc\small  Result in $10^{-11}$ units } 
\\ \hline     
QED (leptons) & $11~6584~718.09\pm 0.14\pm 0.04_{\alpha}$ \\
HVP(lo) & 
$6~908 \pm
39_{\mbox{\rm
\tiny exp}}
\pm 19_{\mbox{\rm
\tiny rad}}\pm 7_{\mbox{\rm
\tiny pQCD}}$\\
HVP(ho) &  $-97.9\pm 0.9_{\mbox{\rm \tiny exp}}\pm  0.3_{\mbox{\rm \tiny rad}}$ \\
HLxL & $105\pm 26$ \\
EW & $152\pm 2 \pm 1$ \\ 
 \hline
 Total SM & $116~591~785 \pm 51$\\
  \hline\hline\ 
\end{tabular}}
\end{center}
\vspace*{-1cm}
\end{table}  

\end{document}